\documentclass[prl,twocolumn,amsmath,amssymb]{revtex4}
\usepackage{graphicx}
\usepackage{color}
\usepackage{graphicx}
\usepackage{dcolumn}
\usepackage{bm}
\usepackage{amsmath}
\usepackage{amsfonts}
\usepackage{bbm}
\usepackage{subfigure}
\usepackage{setspace}

\usepackage{pstricks}
\usepackage{color}
\usepackage{eufrak}
\usepackage{pxfonts}
\newcommand{\beq}{\begin{eqnarray}}
\newcommand{\eeq}{\end{eqnarray}}

\newcommand{\bmp}{\noindent\begin{minipage}{16cm}}
\newcommand{\emp}{\end{minipage}\vskip 7mm} 

\def\drawbox#1#2{\hrule height#2pt
        \hbox{\vrule width#2pt height#1pt \kern#1pt
              \vrule width#2pt}
              \hrule height#2pt}

\def\Asym#1#2{\vcenter{\vbox{\drawbox{#1}{#2}
              \kern-#2pt 
              \drawbox{#1}{#2}}}}

\begin{document}
\title{\Large  \color{red} iT I M P \\ ~\\ isotriplet  Technicolor Interacting Massive Particle as Dark Matter}
	\author{$^{\color{blue}{\spadesuit}}$Mads T. {\sc Frandsen}}
\email{m.frandsen1@physics.ox.ac.uk}
\author{$^{\color{blue}{\varheartsuit}}$Francesco {\sc Sannino}}
\email{sannino@cp3.sdu.edu}
\affiliation{$^{\color{blue}{\spadesuit}}$ {Rudolf Peierls Centre for Theoretical
Physics,
University of Oxford,
1 Keble Road,  Oxford OX1 3NP, United Kingdom.}\\
$^{\color{blue}{\varheartsuit}}$  CP$^\mathbf 3$-Origins, Southern University of Denmark, Campusvej 55, DK-5230
Odense M, Denmark}
\begin{abstract}
We suggest that a weak isotriplet composite scalar possessing an unbroken $U(1)$ global symmetry naturally arises in technicolor models leading to an interesting type of dark matter candidate: the iTIMP. We propose explicit models of the iTIMP, study earth based constraints and suggest possible collider signals. 
\end{abstract}
{\it CP$^3$- Origins: 2009-19}


\maketitle

\section{Introducing the \lowercase{i}TIMP}
 Recent progress in the understanding of the phase diagram of asymptotically free gauge theories \cite{Sannino:2004qp,Dietrich:2005jn,Dietrich:2006cm,Ryttov:2007sr,Ryttov:2007cx} has led to renewed interest in models of dynamical electroweak symmetry breaking \cite{Weinberg:1979bn}. {}For a recent review of the latest developments see \cite{Sannino:2009za}  while an earlier review is \cite{Hill:2002ap}.  {}Explicit examples of technicolor models, not in conflict with electroweak precision tests, have been put forward in \cite{Sannino:2004qp,Dietrich:2005jn,Dietrich:2006cm,Foadi:2007ue,Ryttov:2008xe}. The simplest incarnations of these models are known as (Ultra) Minimal Walking Technicolor models \cite{Sannino:2004qp,Dietrich:2005jn,Dietrich:2006cm,Ryttov:2008xe} and indicated in short by MWT and UMT respectively. The principal feature is that the gauge dynamics is such that one achieves (near) conformal dynamics for a small number of flavors and colors. 
Cold dark matter candidates can be constructed via either the lightest technibaryon, here termed Technicolor Interacting Massive Particles (TIMP)s \cite{Nussinov:1985xr,Gudnason:2006ug,Ryttov:2008xe,Foadi:2008qv}, or new heavy leptons naturally associated to the technicolor theory  \cite{Kainulainen:2006wq}. The TIMP is naturally of asymmetric dark matter type \cite{Nussinov:1985xr}, meaning that its relic density does not have a thermal origin. Within the (U)MWT models such a relic density has been estimated in \cite{Gudnason:2006ug,Ryttov:2008xe} and the results can be used for the dark matter envisioned below.
For models addressing explicitly the issues of extended technicolor interactions (ETC) \cite{Dimopoulos:1979es}, some of which include TIMP type dark matter candidates, see e.g. \cite{Appelquist:2002me}. These models require information on the phase diagram of chiral gauge theories \cite{Sannino:2009za}, see also \cite{Poppitz:2009uq}. 
\newline\indent
Here we suggest that some of the most phenomenologically interesting candidates are electroweak charged technibaryons. These are identified with the lightest of the weak isotriplet technibaryon states which we term iTIMP. They can be pseudo-Goldstone bosons and thus naturally light with respect to the TeV scale. The presence of generic isospin splitting dramatically reduces the cross section of the iTIMP with ordinary nuclei making them phenomenologically viable cold dark matter candidates. 

The iTIMP is the neutral isospin zero component of a weak complex triplet with the following electric charges  
\beq
\label{isodm}
T^+ \ , \quad T^0 \ , \quad T^- \ ,
\eeq
possessing also an extra $U(1)_{TB}$ technibaryon number. We require that $I^3(T^0) = 0$ with $I^3$ the weak isospin operator and assume that $T^0$ is the lightest state in the triplet due to isospin interactions raising the masses of $T^\pm$. 
This is true for the ordinary pions in QCD, where the mass splittings are of the order of
$m_{\pi^\pm}-m_{\pi^0}\sim 5 \ {\rm MeV} $.
By scaling up this mass splitting to the electroweak scale we estimate: 
$\Delta M \equiv M_{T^\pm}- M_{T^0} \sim O(10) \ {\rm GeV}$.
A more precise estimate is \cite{Dietrich:2009ix}:
$\displaystyle{\Delta M^2 \equiv M_{T^\pm}^2- M_{T^0}^2 = 2\,M_{T^{\pm}}^2 \frac{x^2}{1+x^2} }$,
with $x$ the ratio of the weak to hypercharge coupling $x=\frac{g'}{g}$. A typical phenomenologically acceptable value for the mass of these pseudo-Goldstone bosons is around a few hundred GeV leading to a splitting in rough agreement with the estimate above. 
Since $T^0$ is a neutral isospin zero state it will have direct charged current interactions but no neutral ones. The interaction terms will depend on how the iTIMP is built in terms of the elementary techniquarks. {}For example in MWT, which works as a general template, the direct charged current interactions read: 
\beq
\label{CC int}
\mathcal{L}_{CC}&=& i \,\frac{g}{2} (T^{0*}\overleftrightarrow{\partial_\mu}T^{-}W^{+\mu} + T^{0*}\overleftrightarrow{\partial_\mu}T^{+}W^{-\mu}) + h.c.
\eeq
\section{Direct Detection and LHC Phenomenology}
At the tree level, by construction, the iTIMP interacts inelastically with nuclei via W exchange. Since the mass splitting naturally satisfies the inequality (see e.g. \cite{MarchRussell:2008dy}):  
\beq
\Delta M \gg \frac{\beta^2 M_N M_{T^0}}{2(M_N+M_{T^0})} \sim O({\rm keV}) \ ,
\eeq
 with $M_N$ the mass of the target nucleus and $\beta=\frac{v_\textrm{rel}}{c}$ the relative velocity of the iTIMP, this interaction is not relevant here. However, dynamical models realizing a much smaller splitting of $O(10)$ MeV could in principle account for the DAMA/LIBRA signal, along the lines of 
\cite{Bai:2009cd}. {}For example the 'un-technibaryons' arising if the underlying gauge theory is truly conformal as in \cite{Sannino:2008nv}, 
naturally can lead to splittings of $O(100)$ MeV or less. One loop corrections introduce a spin independent cross section which, in the case of a non-composite scalar triplet, has been evaluated in \cite{Cirelli:2005uq} 
 \begin{equation}
 \sigma_{\textrm{nucleon}}^{W} \sim \frac{\pi \, \alpha_2^4 m_N^4}{M_W^2}   \left(\frac{1}{M_W^2}+\frac{1}{M_H^2}\right)^2
 \end{equation}
 with $M_H$ the mass of the (composite) Higgs, $M_W$ the $W$ boson mass and $\alpha_2 = g^2/4\pi$ . 

\begin{figure}[htp!]
{\includegraphics[height=5cm,width=8.5cm]{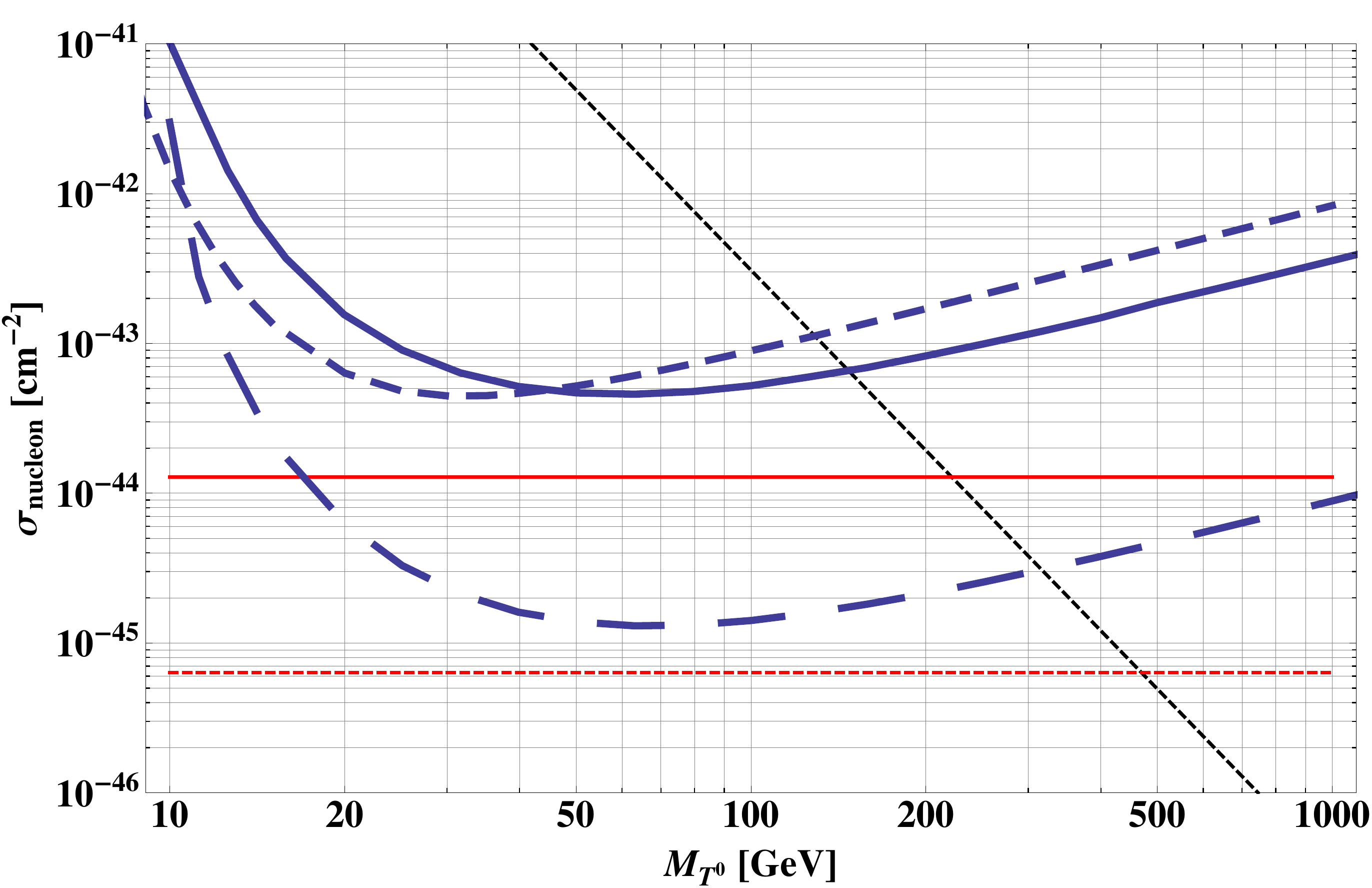}}
\caption{ TIMP -- nucleon cross section: Black dashed line corresponds to the composite Higgs exchange with $d_M=1, f=0.3$ and $M_H=300$ GeV. Solid and dashed horizontal red lines correspond to the photon and 1-loop W exchange cross sections with $d_B =1$ and $\Lambda =3$ TeV. Also plotted are the exclusion limits from CDMS II Ge Combined  (solid-thick-blue), XENON10 2007 (dashed-thick-blue) and projected SuperCDMS.}\label{itimpexclusion}
\end{figure}

Since the iTIMP is a composite particle with electroweakly charged constituents it will, in general, have an effective coupling to the photon, corresponding to a charge radius operator \cite{Bagnasco:1993st}, as well as a coupling to the composite Higgs \cite{Foadi:2008qv}. The latter contribution arises in generic models of scalar dark matter as well \cite{McDonald:1993ex}. The associated independent cross sections with nuclei are: 
\beq
\sigma_{\textrm{p}}^{\gamma} &=& \frac{\mu^2}{4\pi} \left[\frac{ 8\pi \, \alpha \, d_B}{\Lambda^2}\right]^2 \ , \quad
\sigma_{\textrm{nucleon}}^H = \frac{\mu^2}{4\pi}   \left[\frac{d_M  \, f m_N}{M^2_H M_{T^0}}\right]^2  ,
\eeq
where $d_B , d_M$ are $O(1)$ couplings, $\Lambda$ is the compositeness scale, $\mu$ the nucleon-iTIMP reduced mass and $f$ parameterizes the composite Higgs-nucleon coupling. We refer to \cite{Ohki:2008ff} for recent results on the strange quark contribution to $f$ which we fix to be $f=0.3$ for a qualitative estimate.
We plot the cross-sections against the exclusion limits from CDMS and XENON \cite{Ahmed:2008eu} in Fig.~\ref{itimpexclusion}. 
We find that for masses $M_{T^0} \gtrsim 100$~GeV, $M_H \gtrsim 300$~GeV,  and a natural compositeness scale of $\Lambda \gtrsim 3$~TeV  the iTIMP would have escaped detection.

At the Large Hadron Collider (LHC) the iTIMP may be pair produced via 
decays of the composite Higgs $H$ or $T^\pm$.  In \cite{Foadi:2008qv} the first of 
these processes was studied  assuming that $H$ was produced in association with a SM gauge boson. 
 It was found to be a promising channel at the LHC, 
yielding a signal with two leptons and large missing transverse momentum.
Here we consider the second process, shown in Fig.~\ref{iTIMP 
production} featuring either a $W$ or $Z$ boson in the intermediate state. Interestingly the second diagram is of the 'antler' type \cite{Han:2009ss} which might allow determination of the masses of both $T^{\pm}$ and $T^0$ at the LHC.
\begin{figure}
{\includegraphics[height=2cm,width=8cm]{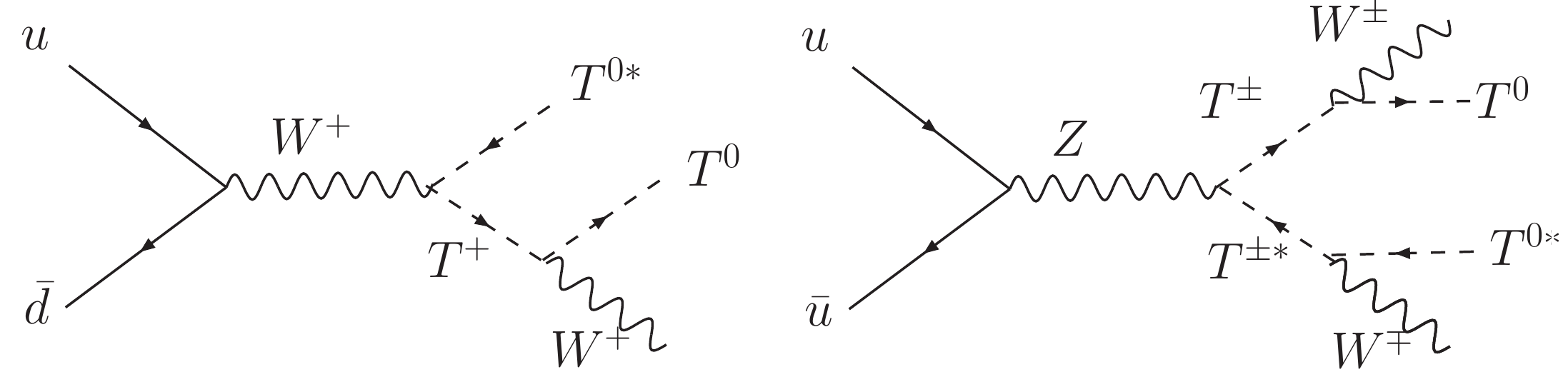}}
\caption{iTIMP production at the LHC via $W$ or $Z$ exchange.}\label{iTIMP production}
\end{figure}
\begin{figure}
{\includegraphics[height=5cm,width=8.5cm,clip]{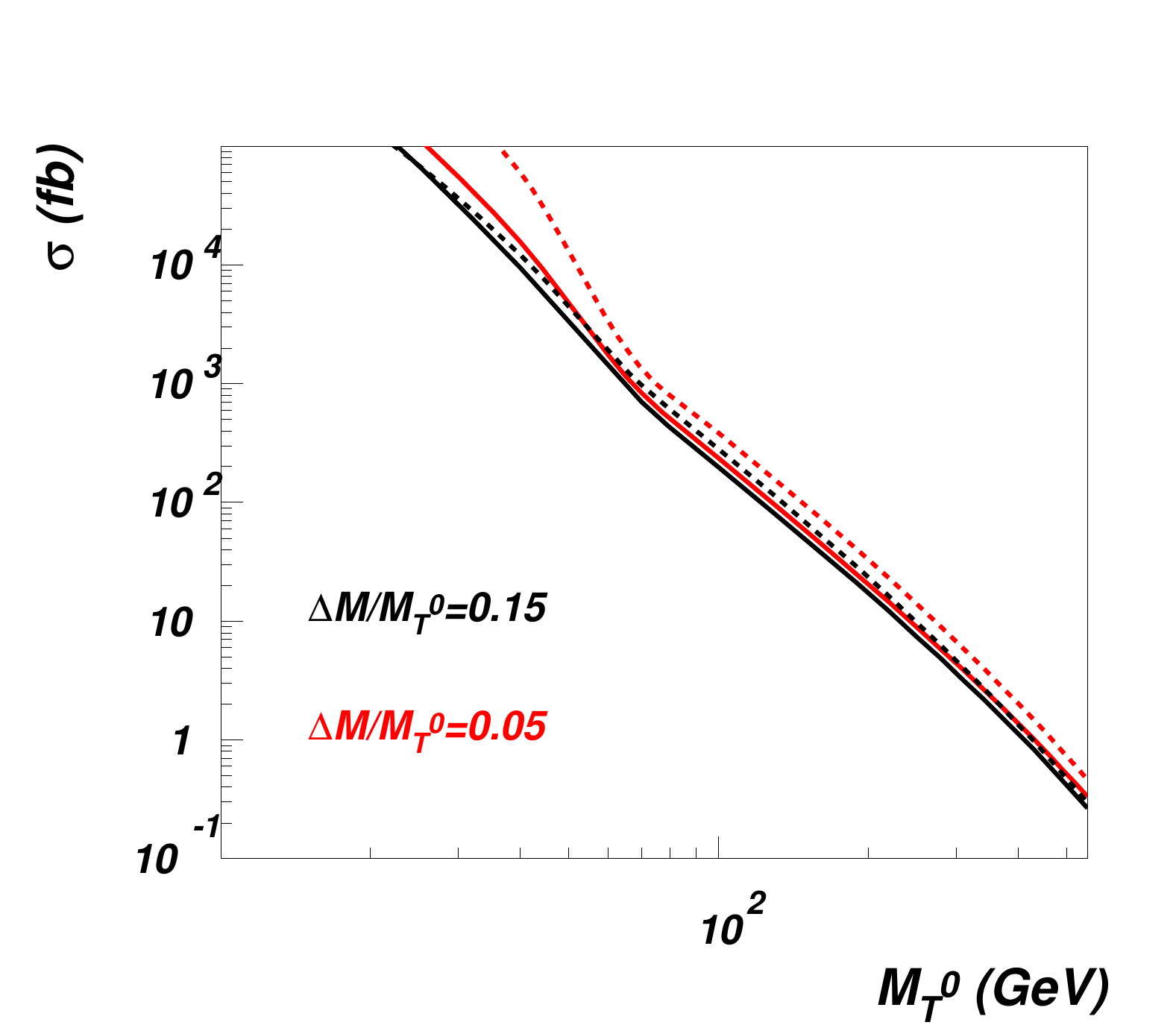}}
\caption{Production cross section of $T^0 T^{0\ast} W^+$ (solid) and $T^0 T^{0\ast} W^+ W^-$at 
the LHC as a function of $M_{T^0}$ for two values of the isotriplet mass 
splitting $\Delta M$.}\label{iTIMP prod cs}
\end{figure}
As shown in Fig.~\ref{iTIMP prod cs}, for $M_{T^0} < 250$ GeV we observe a 
production cross-section of at least 10 fb. 
 We expect the effects of the spin one massive states present in a generic technicolor model (if light) to enhance the signals as in \cite{Foadi:2008qv} and note that similar processes have been studied within MWT where the role of the iTIMP is played by new heavy leptons \cite{Frandsen:2009fs}.

\section{Models for the \lowercase{i}TIMP}
Before providing models for the iTIMP we summarize its salient features: The iTIMP
\newline \noindent
0) is the neutral isospin zero component of a weak isotriplet.
\newline \noindent
i) $U(1)_{TB}$ symmetry is natural. 
\newline \noindent
ii) relic density is related to the baryon one via an asymmetry. 
\newline \noindent
iii) is a composite state which  can be light being a pseudo Goldstone boson.
\newline \noindent
Properties i)-iii) are shared by the TIMP defined in \cite{Ryttov:2008xe,Foadi:2008qv}  while the difference is that the iTIMP is charged under the weak interactions yielding potentially richer  collider phenomenology.

Now, consider a generic technicolor model with only one electroweak doublet of technifermions  belonging to a representation $R$ of the underlying technicolor gauge group. With respect to the electroweak interactions the charge assignment for the technifermions can be:
\beq
\label{basemodel}
 Q_L =& \left(\begin{array}{c} U_L^{+1/2} \\D_L^{-1/2} \end{array}\right) \ , \quad U_R^{+1/2} \
, \ D_R^{-1/2} \ , 
\\
Y(Q_L) =& 0 \ , \quad
Y(U_R,D_R)=\left(1/2, -1/2 \right)
\eeq
The upper index refers to the electric charge of each state. If $R$ is real the flavor symmetry breaking pattern is:
$SU(4)\to SO(4)$. 
This leads to nine Goldstone bosons, three of which are absorbed by the SM gauge bosons. Six additional uneaten Goldstone bosons with technibaryon charge form triplets as in Eq.~(\ref{isodm}). The isospin zero component  is the iTIMP. 

If  $R$ is pseudo-real the flavor symmetry breaking pattern is:
$SU(4)\to Sp(4)$.
This leads to five Goldstone bosons, three of which are absorbed by the SM gauge bosons. The additional two form a complex scalar charged under the technibaryon number, which is the TIMP introduced in \cite{Ryttov:2008xe}.

Explicit examples of $SU(N)_{TC}$ technicolor models realizing the iTIMP via the $SU(4)$ breaking to $SO(4)$ are: Theories with one weak doublet of Dirac fermions transforming according to the adjoint of $SU(N)_{TC}$, with $N=2$ being the MWT model \cite{Sannino:2004qp}, and an $SU(4)_{TC}$ gauge theory with one weak doublet of Dirac fermions transforming according to the two-index antisymmetric representation \cite{Dietrich:2006cm}. 

It is useful to show the technibaryon wave-functions, e.g. in the $SU(4)_{TC }$ case we can construct 
 $\epsilon_{c_1c_2 c_3 c_4} Q_{L}^{c_1c_2, f} Q_{L}^{c_3 c_4, f'}  \sigma_{f f'}^a$
with
$a=0,...,3$
where $\sigma^0$ is the identity matrix and $\sigma^{1,2,3}$ the Pauli matrices in flavor space, $c_i = 1,...,4$  (for any $i$) is the technicolor index and $f$ and $f'$ are flavor indices ranging from one to two. 
Suppressing technicolor and spin indices this gives the bilinears  \footnote{The technibaryon wave function presented in \cite{Dietrich:2006cm} for $SU(4)_{TC}$ with technicolor fermions in the two-index antisymmetric representation vanishes identically.   }:
\beq
 Q_L^{f} Q_L^{ f'}  \sigma_{f f'}^1 &\sim &   U_L D_L \ ,
\\ \nonumber
  Q_L^{f} Q_L^{f'}  (\sigma_{f f'}^0+\sigma_{f f'}^3) &\sim &  U_L U_L  \ ,
\\ \nonumber
Q_L^{ f} Q_L^{ f'}  (\sigma_{f f'}^0-\sigma_{f f'}^3) &\sim &   D_L D_L \ .
\eeq
Analogous bilinears may be constructed with the aid of the $\delta$-symbol (in technicolor space) in the   MWT model.

Typically for theories with fermions in the adjoint representation we can construct bound states made out of a technifermion and a technigluon. {}For the hypercharge choice made here these will be fractionally charged states. They need to be either depleted very efficiently during the evolution of the Universe, as suggested in \cite{Goldberg:1982af}, or forbidden via, for example, a new confining symmetry. To avoid this problem we consider instead $SO$ gauge theories for the technicolor gauge dynamics, observing that: i) {}For $SO$ the defining representation is real, leading to the chiral symmetry breaking pattern $SU(4)\to SO(4)$, which yields the iTIMP; ii) The adjoint representation of $SO$ is the two-index antisymmetric and hence one cannot form gauge-singlet states made out of one technigluon and one technifermion for a number of technicolors larger than three. For the $Sp$ case we can form, instead, the TIMP and the adjoint representation corresponds to the two-index symmetric one again forbidding the formation of fractionally charged technicolor singlets. 

\begin{figure}[h!]
{\includegraphics[height=5cm,width=7.5cm]{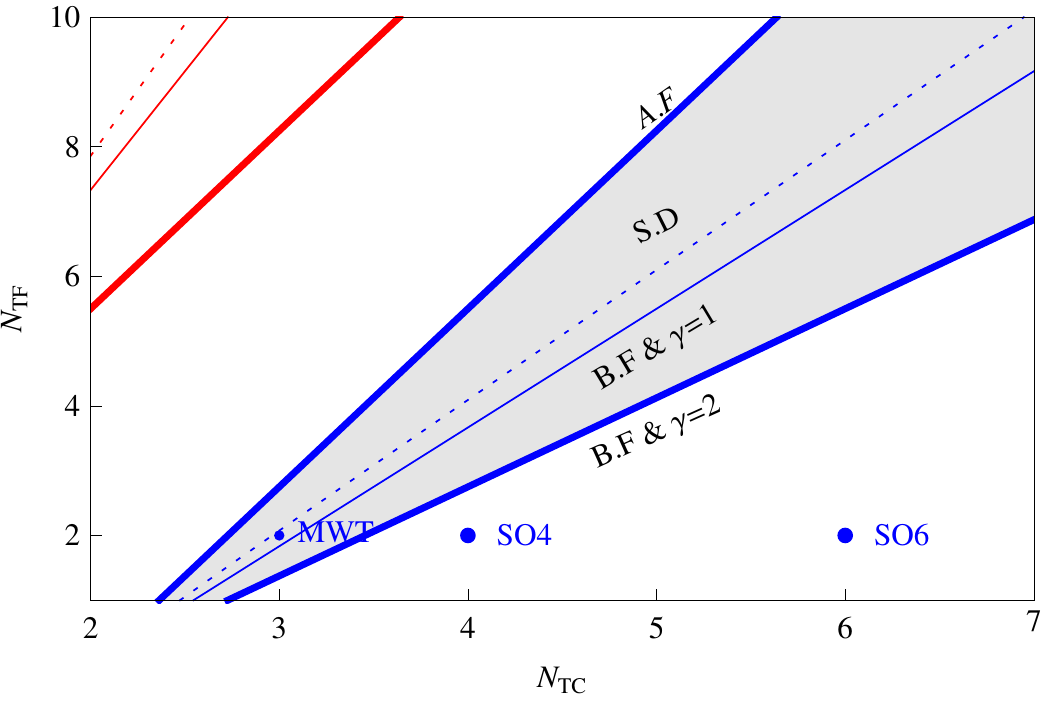}}
\caption{Conformal phase diagram of $SO(N)$ (blue) and $SU(N)$ (red) gauge theories with $N_f$ Dirac flavors in the vector representation. The upper thick line (A.F) is the loss of asymptotic freedom, only visible for $SO(N)$. The lower solid lines are the lower bound on the conforwal window using the conjectured all orders beta function asuming the anomalous mass dimension of the fermion bilinear is $\gamma=2,1$ respectively while the dotted line uses ladder approximation to the Schwinger-Dyson equation (S.D) \cite{Appelquist:1988yc}. The MWT model and the new models considered here are indicated in red.}\label{SO4MWT}
\end{figure}

The phase diagram of $SO$ and $Sp$ theories have been explored in \cite{Sannino:2009aw}. In Fig.~\ref{SO4MWT} we compare $SO$ (blue) to $SU$ (red) theories with $N_f$ Dirac fermions in the vector representation.
The upper thick line (A.F) indicates loss of asymptotic freedom. The lower solid lines are the lower bound on the conformal window using the conjectured all orders beta function \cite{Ryttov:2007cx}
asuming the anomalous mass dimension of the fermion bilinear is $\gamma=2,1$ respectively, while the dotted line uses the ladder approximation to the Schwinger-Dyson equation (S.D) \cite{Appelquist:1988yc}. We wish to identify possible near conformal gauge theories minimizing the contributions to the electroweak precision tests while possessing interesting iTIMP candidates. {}The near conformal nature of the theory  reduces the tension with precision observables as suggested in 
\cite{Appelquist:1998xf}. To get a qualitative estimate of the $S$ parameter we use its naive expression obtained via one-loop of massive technifermions. {}For $SO(N)$ theories with 2 Dirac fermions in the vector representation this is $S = \frac{1}{6\pi} N$. The MWT model is equivalent to the $SO(3)$ theory with $S\approx 1/(2\pi)$ while $SU(3)_{TC}$ with adjoint technifermions gives $4/(3\pi )$. {}The $SU(4)_{TC}$ model with technifermions in the two-index antisymmetric representation is equivalent to the $SO(6)$ theory which gives $1/\pi$.  There are strong indications, from analytic \cite{Sannino:2004qp,Dietrich:2006cm,Ryttov:2007cx}, as well as lattice studies \cite{Catterall:2007yx} 
that the MWT model is indeed (near) conformal.  
Based on this we propose a new minimal candidate walking technicolor model including in its spectrum the iTIMP dark matter candidate: The $SO(4)$ gauge theory with 2 Dirac flavors (equivalently 4 Weyl flavors) in the vector representation which we term the OMT model. The naive $S$ parameter is $\frac{2}{3\pi}$ and the model appears safely outside the conformal window using both the Schwinger-Dyson and the conjectured all orders beta function. However, if the MWT model is truly conformal \cite{Catterall:2007yx} then by inspecting the phase diagram, the OMT model might be walking.  
\newline \indent 
We proposed the iTIMP as a viable new type of technibaryon dark matter and suggested several new explicit model realizations. We considered the direct detection contraints and collider phenomenology.


\end{document}